%% file: Dispersion.tex
\documentclass[%
 reprint,
 amsmath,amssymb,
 aps,
 pra
]{revtex4-1}

\usepackage{graphicx}
\usepackage{dcolumn}
\usepackage{bm}
\usepackage{comment}
\usepackage{todonotes} 

\usepackage[hidelinks]{hyperref}
\hypersetup{
    colorlinks=true,
     linkcolor=blue,
     filecolor=blue,
     citecolor = blue,      
     urlcolor=black,
}

\begin{document}

\preprint{APS/123-emitterD}

\title{Green's functions of and emission into discrete anisotropic and hyperbolic baths}%

\author{Lewis Ruks}
\email[]{lewis.ruks@oist.jp}
\author{Thomas Busch}%
\affiliation{\mbox{Quantum Systems Unit, Okinawa Institute of Science and Technology Graduate University,}\\ Onna, Okinawa 904-0495, Japan
}

\date{\today}

\begin{abstract}
    In this work, we study wave propagation in generic Hermitian local periodic baths, and investigate the effects of anisotropy and quasi-breaking of periodicity on resonant emission into the band of the bath. We asymptotically decompose the Green’s function into long-range travelling waves composed of all wavevectors (near-)resonant at the emitter frequency, and rapidly decaying evanescent waves. Our approximation then converges exponentially with increasing source-receiver separation $\bm{\rho}$ when resonant wavepackets with group velocity parallel to $\bm{\rho}$ exist. In hyperbolic media this condition may not be satisfied, and we find that the exponential decay length of oscillating evanescent waves locally around caustics generally depends as a power law with exponent 3/2 on the angle made between $\bm{\rho}$ and the caustic. For $\bm{\rho}$ beyond the caustic we observe that the Green's function can become almost imaginary, which results in exclusively incoherent emitter-emitter interactions and allows the simulation of purely dissipative systems with short-range interactions. Here the interaction length is tunable via the separation vector of the emitters. We finally probe the hyperbolic dispersion beyond the previous regimes by applying an artificial gauge field on the lattice. We find that emission resonant with the corresponding open orbits in the Brillouin zone is quasi-one dimensional, in contrast to an isotropic environment. The quasi-1D emission is further topologically protected against local and global lattice perturbations and periodically refocussing, offering a robust bi-directional transport of excitations in higher-dimensional media.  
\end{abstract}

\maketitle
\section{Introduction}
The paradigmatic setting underpinning (quantum) optics is that of emitters coupled to a (photonic) bath. Since the discovery of the Purcell effect led to the general realization that modifying the bosonic mode structure through a cavity effects the emission profile in the weak-coupling regime \cite{purcell_effect}, structured baths have received a great deal of attention from both the classical \cite{crystal_attention,acoustic_attention,slow_light_attention} and quantum \cite{rao_longrange,rabi_attention,optical_switch_attention,reservoir_anisotropic,reservoir_decay,reservoir_isotropic,reservoir_review} communities. Spurred on by recent technological advances in coupling at the micro- and nano-scale \cite{dom_qe_coupling,qe_cool,two_dim_cool_trap,two_d_emission_possible,embedded_emitter,evidence_Super}, many theoretical proposals for realising exotic phenomena by coupling emitters to structured baths have been discussed particularly in the last decade~\cite{lukin_green_3,hood_green,tunable_tudela,scatter_weyl,waveguide_review,cirac_noperturb2017,extended_dipole,two_array_band}, offering potential applications in quantum information~\cite{waveguide_entanglement,waveguide_entanglement_two,quantum_computation} and simulation of many-body physics~\cite{scully_super,band_scaling,photon_polariton,quantum_chem,top_bath}. Central to all dynamics is the dispersion relation and dimensionality of the underlying bath, which determines the spatial emission profile of a weakly coupled emitter. \\
Generically, resonant emission into a band of a periodic bath is phenomenologically understood to consist of a travelling part and a rapidly  decaying evanescent part~\cite{extended_dipole}, whose natures are determined by the geometry of the bath dispersion relation at the resonant frequency~\cite{scatter_weyl,geometry_van_hove,tudela_dirac,tunable_tudela}. Whilst transport in general 1D baths can often be exactly quantified using complex analysis~\cite{fourier_decay,hood_green,rao_longrange,wannier_decay}, a concrete characterisation of the travelling part and accompanying phenomena in generic environments of dimension $d>1$ is currently restricted to lattice-specific analyses~\cite{extended_dipole,top_bath,embedded_emitter,cirac_noperturb2017}. In addition, dimensions $d>1$ allow for strongly anisotropic propagation of waves in hyperbolic media and whilst materials exhibiting hyperbolic dispersion realize exotic wave propagation mediating novel emitter-emitter interactions~\cite{hyperbolic_disp,hyperbolic,directional_gap,poddubny_hyperbolic}, studies are often made in the continuum medium approximation \cite{hyperbolic_disp_discrete} and focus on the emission profile within caustics. The effect of artificial gauge fields on the emission into hyperbolic lattices also remains unclear, despite their experimental realisation and investigations in isotropic media~\cite{artificial_gauge,semiclassical_electron,photon_polariton}.\\
In this work, we investigate emission into general local periodic baths with the specific example of the square tight-binding lattice, using the bath Green's function and exact simulations. We use differential geometry to asymptotically derive the Green's function of a generic local $d$ dimensional ($d$D) bath as an integral over the resonant level set that converges up to exponential errors due to evanescent waves. This result offers a cheap and accurate alternative to the stationary phase method~\cite{stationary_optics,phonon_focus,photon_focus} to calculate the Green's function for large source-receiver separations $\bm{\rho}$. When the emitter is tuned to a frequency in the band exhibiting hyperbolic dispersion, oscillating evanescent waves~\cite{ghost_phase,ghost_wave,directional_gap,ghost_polariton} dominate the Green's function for $\bm{\rho}$ beyond a caustic, and our approximation is no longer valid for these directions. We then investigate this regime, and derive for $\bm{\rho}$ almost parallel to the caustic a universal scaling of the decay length of oscillating evanescent waves with respect to the off-caustic angle. We also find that emitter-emitter exchanges effected through the bath may be dominantly incoherent for almost all directions beyond the caustic. In conjunction with the previous finding, this mechanism allows for simulation of purely dissipative nearest-neighbor spin models, where the interaction length is tunable within a band and via the separation direction between the emitters, as opposed to, e.g., detuning from (beyond) a band edge~\cite{band_scaling,two_array_band,hood_green,directional_gap}. We finally probe the associated open orbits~\cite{open_orbit} in quasimomentum space via an effective magnetic field on the lattice. Quasi-1D emission with non-zero transport may be realised, in contrast to isotropic magnetic orbits~\cite{photon_polariton} associated with Landau orbitals. Simulating exactly, we find the emission to periodically and robustly refocus down to the single site level in the plane transverse to propagation. Consequently, emitters located at the refocussing points strongly interact to permit high-fidelity bound states in the continuum \cite{bound_continuum,chang_cavity,sinha_nonmarkov} using only a few emitters. The topological protection of open orbits enables preservation of transport in the presence of local obstructions and moderate global disorder. This final result provides bidirectional transport and storage of excitations in higher dimensional media.\\

\section{formalism and setup}

\begin{figure*}[t!]
    \centering
    \makebox[0pt]{\includegraphics[scale=0.40]{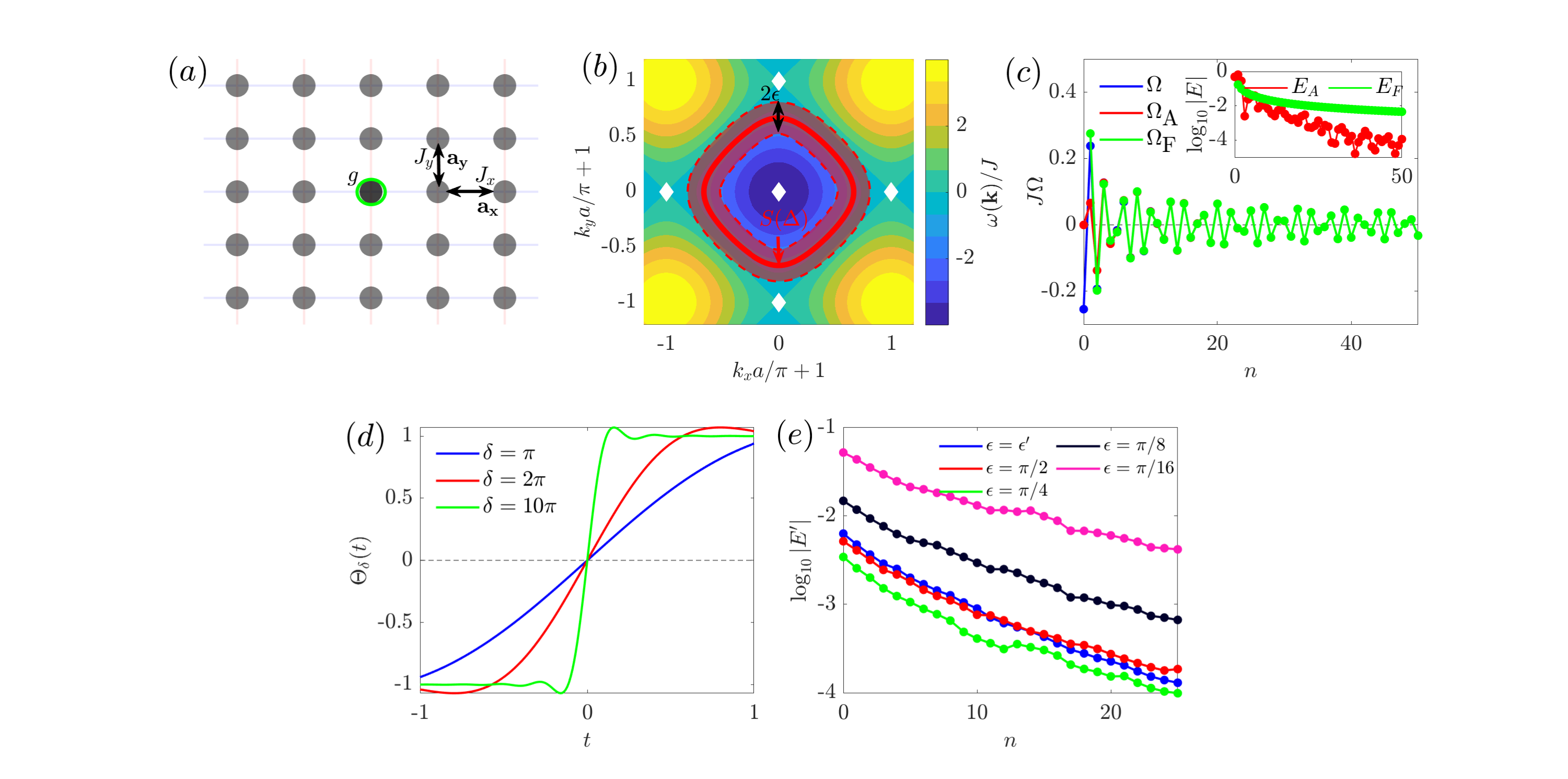}}
    \caption{(a) Schematic of the tight binding lattice, with the emitter denoted as a green circle. (b) Dispersion relation of the isotropic square tight-binding lattice. White diamonds indicate $v(\mathbf{k}) = 0$, whilst red highlights the tube of radius $\epsilon$ about $S$, bordered by the dashed lines.  (c) $J\Omega$ computed exactly as in Appendix ~\ref{app:numerics} together with the approximations (A) given through Equation \eqref{eq:medrange} and (F) through equation Equation \eqref{eq:longrange} as $\bm{\rho} = n(\mathbf{a}_{x} + \mathbf{a}_{y})$ is varied with $n$. We here take $\epsilon = \epsilon' = 0.675.$ Inset: relative error $E = |\Omega - \Omega_{i}|/|G|,$ ($i=A,F$) on a log scale. (d) The phase amplitude function~\eqref{eq:phase_amp} plotted for varying $\delta = \rho\epsilon$. (e) 26-point moving average of the logarithm of the error as tube radius $\epsilon$ varies (for $a=1$). We accordingly have $E'(n) = (\prod_{i=0}^{25}E(n+i))^{1/26}$.}\label{fig:initialcompare}
\end{figure*}

We consider a local (i.e., coupling to only finitely many neighbors) and periodic $d$-dimensional tight-binding Hamiltonian $\hat{H}$, which describes a lattice with the lattice sites given by $\mathbf{r}_{i}$. The Hamiltonian is linear in annihilation, $\hat{a}_{\mathbf{r}_{i}}$, and creation, $\hat{a}_{\mathbf{r}_{i}}^\dagger$, operators for bosonic excitations at site $\mathbf{r}_{i}$, so that (with $\hbar = 1$) $\hat{H} = \sum_{\mathbf{r}_{i}\mathbf{r}_{j}}J_{\mathbf{r}_{i}\mathbf{r}_{j}}\hat{a}_{\mathbf{r}_{i}}^{\dagger}\hat{a}_{\mathbf{r}_{j}}$ with $[\hat{a}_{\mathbf{r}_{i}},\hat{a}_{\mathbf{r}_{j}}^{\dagger}] = \delta_{\mathbf{r}_{i}\mathbf{r}_{j}}$. Propagation in periodically structured baths can be analysed through the \textit{Green's function} $\hat{{G}}(\Delta) = (\Delta-\hat{H})^{-1}$, which has off-diagonal elements ${G}(\mathbf{r},\mathbf{r}',\Delta) = \langle \mathbf{r} | \hat{G} | \mathbf{r}'\rangle$, where translational invariance means that these only depend on the separation $\bm{\rho} = \mathbf{r} - \mathbf{r}'.$
When $\hat{H}$ is Hermitian, the Sokhotski-Plemelj theorem permits a decomposition of $G$ into coherent and incoherent terms as $G(\mathbf{r},\mathbf{r}',\Delta) = G = {\Omega} - \frac{i}{2}{\Gamma}$ with (assuming a single band)
\begin{align}
		\label{eq:gam}
		{\Gamma} &= -2\mathfrak{I}[G] = \mathcal{A} \int_{S}\frac{d^{d-1}\mathbf{k}}{(2\pi )^{d-1}}\frac{\psi(\mathbf{k},\mathbf{r})\psi^{*}(\mathbf{k},\mathbf{r'})}{{v}(\mathbf{k})}, \\
		\label{eq:om}
		{\Omega} &= \mathfrak{R}[G] = \mathcal{A}\mathcal{P}\int_{\text{BZ}}\frac{d^{d}\mathbf{k}}{(2\pi )^d}\frac{\psi(\mathbf{k},\mathbf{r})\psi^{*}(\mathbf{k},\mathbf{r'})}{\Delta-\omega(\mathbf{k})},
\end{align}
where $\mathcal{A}$ is the volume of the unit cell, $\omega(\mathbf{k})$ is the band dispersion, and $\psi(\mathbf{k},\mathbf{r}) = \langle \mathbf{r} | \psi(\mathbf{k}) \rangle$ is the Bloch wavefunction. The coherent part $\Omega$ comprises a singular principal value integral $\mathcal{P}\int$ over the Brillouin zone (BZ), whilst the dissipative part $\Gamma$ sums over the \textit{resonant level set} $S$ with $S = S(\Delta) = \{\mathbf{k} : \omega(\mathbf{k}) = \Delta\},$ weighted by the group velocity $v = |\mathbf{v}(\mathbf{k})| = |\nabla \omega(\mathbf{k})|$. In artifical light-matter systems $S$ is highly controllable due to the tunability of  $\Delta$ through the emitters~\cite{scatter_weyl,two_array_band,hood_green,lukin_green_3,band_scaling,quantum_chem}, and can be probed using the same emitters. We will consider spin-$\frac{1}{2}$ emitters (although the following results are generally applicable to classical dipolar emission) of resonant frequency $\Delta,$ each coupled to a single site $\mathbf{r}, \mathbf{r}'$ of the bath with strength $g$. In the thermodynamic bath and weak coupling limits, one can make the Born-Markov approximation to obtain the effective dipole-dipole coupling elements in the standard master equation as \cite{zoller_Markov,tunable_tudela} 
\begin{equation}
\label{eq:spincouplings}
V_{\mathbf{r}'\mathbf{r}'} = g^2 G(\mathbf{r}',\mathbf{r}',\Delta).
\end{equation}
Up to scaling, the emitter dynamics are therefore entirely determined through $G,$ and chiefly through $\omega(\mathbf{k})$ and $\Delta.$ Whilst we present our results in the context of emitters coupled to tight-binding lattices, we note that qualitative phenomena are observable \cite{wave_propa} and systematically reproducible \cite{latticecontinuum} in continuum media, including ultracold gases \cite{gas_lattice,quantum_chem}, photonic crystals \cite{binding_photonic,tight_photonic,crystal_lattice_bic}, and acoustic meta-crystals \cite{acoustic_crystal,acoustic_crystal_two}. In particular we emphasize that our investigation applies generally in linear and periodic media. To concretely illustrate results in the following we consider the tight-binding anisotropic square lattice shown in Fig.~\ref{fig:initialcompare}(a). The nearest neighbour coupling strength is given by $J_{x(y)}$ for coupling in the $x (y)$ direction along lattice vectors $\mathbf{a}_{x} = a\hat{\mathbf{x}}$ $(\mathbf{a}_{y} = a\hat{\mathbf{y}})$. The dispersion is then given by $\omega(\mathbf{k}) = 2(J_{x}\cos(k_{x}a) + J_{y}\cos(k_{y}a))$, and we set the emitter to be located at the origin $\mathbf{0} = (0,0)$.
\label{sec:asymptotic_decomposition}
 \\

\section{Asymptotic decomposition }

The coherent interactions described by Eq.~\eqref{eq:om} are given through a strongly singular continuum sum that is particularly unwieldy beyond one dimension~\cite{hood_green,sanchez_limits}.  Standard methods of complex analysis used in 1D~\cite{fourier_decay,hood_green,rao_longrange} or employed with Wannier functions~\cite{wannier_decay,fourier_decay} can not be readily applied to the strongly singular manifold in Eq.~\eqref{eq:om} when $d>1$. However, $S$ is generically a (union of) smooth $(d-1)D$ dimension manifold(s)~\cite{fermi_surface_1}, and results of differential geometry and asymptotic analysis can be employed for large $\bm{\rho}$. We localize the contributions to $\Omega$ near to and away from $S,$ where the integrand is strongly singular and analytic, respectively, as shown in Fig.~\ref{fig:initialcompare}(b) (in Appendix~\ref{app:long-range:nodegen} we show that for multiple bands the total integrand remains analytic at degeneracies despite the non analyticity of a single band there). Applying Weyl's tube formula~\cite{WeylVolume2017,gray_tubes,tube_backup,tube_backup_two}, we asymptotically obtain (see Appendix~\ref{app:long-range:coherent-part}) the dominant contributions to $G$ from a tubular region of $S$ up to corrections due to evanescent waves that decay exponentially as $\rho = |\bm{\rho}| \to \infty$:
\begin{equation}
\label{eq:medrange}
G \sim -i\mathcal{A} \int_{S}\frac{d^{d-1}\mathbf{k}}{(2\pi )^{d-1}}\Pi_{\rho}(\hat{\mathbf{v}} \cdot \hat{\bm{\rho}} )\frac{\psi(\mathbf{k},\mathbf{r})\psi^{*}(\mathbf{k},\mathbf{r'})}{{v}(\mathbf{k})}.
\end{equation}

Here $\hat{\mathbf{v}} = \mathbf{v}/v$ and $\hat{\bm{\rho}} = \bm{\rho}/\rho$ are the normalized group velocity and separation directions respectively at $\mathbf{k} \in S$, while $\Pi_{\rho}$ is an amplitude function
\begin{eqnarray}
    \label{eq:full_amp}&\Pi_{\rho}(\hat{\mathbf{v}} \cdot \hat{\bm{\rho}}) = \frac{1 + \Theta_{\rho}(\hat{\mathbf{v}} \cdot \hat{\bm{\rho}})}{2} \nonumber \\ 
	\label{eq:phase_amp}&\Theta_{\rho}(\hat{\mathbf{v}} \cdot \hat{\bm{\rho}} ) = \int_{-\infty}^{\infty} \frac{dy \sin((\hat{\mathbf{v}} \cdot \hat{\bm{\rho}})y)\phi_{\rho \epsilon}(y)}{\pi y}
\end{eqnarray}
for the bump function
\begin{equation}
    \phi_{\delta}(y) = \begin{cases} 
      \exp(\frac{y^2}{y^2-\delta^2}) & |y|  < \delta \\
      0 & \text{Otherwise}.
   \end{cases}
\end{equation}
This expression for the Green's function is the first key result of this work. Note the implicit dependence on the parameter $\epsilon$ contributing to the width $\delta = \rho \epsilon$ of the bump function. Here $\epsilon$ is the radius of the tubular region in quasimomentum space (Figure \ref{fig:initialcompare}(b)) and can be chosen within any range such that the tubular region does not self-intersect or cross points of zero group velocity. The behaviour of $\Theta$ is shown in Figure \ref{fig:initialcompare}(d), where as $\hat{\mathbf{v}}$ becomes (anti-)aligned with $\bm{\rho},$ the phase function $\Theta_{\rho}$ tends towards 1 (-1). This behaviour becomes more pronounced for increasing $\rho$ or $\epsilon.$ $\Pi_{\rho}$ then smoothly varies from 0 to 1 as the angle $\hat{\mathbf{v}} \cdot \hat{\bm{\rho}} $ varies, so that $\Pi_{\rho}(t) \to H(t)$ as $\rho \to \infty$, where $H$ is the Heaviside function. With the decomposition $\Pi_{\rho} = \frac{1+\Theta_{\rho}}{2},$ containing the scattered part 1/2 and unscattered part $\Theta_{\rho}/2$ corresponding to $\Gamma$ and $\Omega$ respectively, Equation \eqref{eq:medrange} generalizes the dominant plane wave contribution $e^{ikx}$ from scattering in 1D~\cite{hood_green,fam_omega,pichler_chiralspins,rao_longrange}, and places $\Omega$ on a similar footing to $\Gamma$. For comparison, the usual classical approximation obtained in Appendix~\ref{app:long-range:stationary-phase} by a stationary phase argument (see Equation \eqref{eq:longrange}, \cite{stationary_optics,StatPhase2017,phonon_focus,photon_focus}), is valid up to leading inverse power of $\rho.$ Through Equation \eqref{eq:medrange} power-law decays of all orders are retained for $\Omega$, which can be computed in a similar manner to $\Gamma$ through a regular surface integral. In Figure \ref{fig:initialcompare}(c), the exponential versus power-law decay of the error from approximants \eqref{eq:medrange} and \eqref{eq:longrange} can be seen in the inset. Of course, the convergence of Equation~\eqref{eq:medrange} to $G$ is in general determined by $\epsilon$, and we display this dependence in Figure~\ref{fig:initialcompare}(e). To allow for ease of comparison we plot a moving average of the errors which eliminates the sharp variations over short scales seen in the inset of Figure~\ref{fig:initialcompare}(c). We see that the rate of decay is left unchanged by the choice of $\epsilon$ (within those $\epsilon$ allowed), but an overall discrepancy in leading factor is obtained, so that in practice $\epsilon$ should be chosen to minimize this (i.e., $\epsilon = \pi/4$ in the figure). This variation arises from the fact that we are essentially approximating $G$ as an integral over the tubular region, which is in turn determined by $S(\Delta)$ and $\epsilon.$ Too small $\epsilon$ means that we are neglecting to consider almost all of the BZ in our tubular region, so that a larger $\rho$ is required until the contributions from the remainder become comparatively negligible, whilst too large $\epsilon$ amounts to one approximating more of the BZ by behaviour on $S(\Delta),$ which can lower accuracy when the dispersion and wavefunctions vary significantly within the tubular region~\cite{tunable_tudela,tudela_dirac,lukin_green_1,cirac_noperturb2017}. In general we can expect the rate of variation to remain unchanged with varying $\epsilon$ for a fixed direction, for the complex poles of $\omega(\mathbf{k})$ contribute to the exponentially decaying part of $G$, and all these poles are always neglected by our approximation for any value of $\epsilon$. Then, for a fixed direction of $\bm{\rho},$ $\epsilon$ should in practice be chosen to minimize the leading factor of the error. There is no apparent systematic way to choose the optimal $\epsilon$, but a suitable $\epsilon$ can be deduced by considering the characteristic quasimomentum scale: in Figure~\ref{fig:initialcompare}(b), $\pi/(4a)$ is a suitable candidate for the quasimomentum scale, whilst (for example) $\pi/(16a)$ and $\pi/(2a)$ are too small and too large respectively. This is then reflected in Figure~\ref{fig:initialcompare}(e), where we indeed see that $\pi/4$ (when $a=1)$ outperforms the other two. Further inexpensive optimization could then be performed by sampling errors for small separation vectors. Equation $\eqref{eq:medrange}$ could offer particular utility when strongly subradiant~\cite{subradiant_chain_normal,subradiance_example,exponential_subradiance,subradiant_chain_atypical} dynamics are present in a large system, where additional corrections present in Equation \eqref{eq:om} and beyond those obtained by heuristic arguments~\cite{hood_green,fam_omega,pichler_chiralspins,eldredge_chiralorg} have the potential to observably modify dynamics~\cite{subradiant_chain_atypical} due to $\Omega$ coupling subradiant states.\\

\begin{figure*}[t!]
     \makebox[0pt]{\includegraphics[scale=0.42]{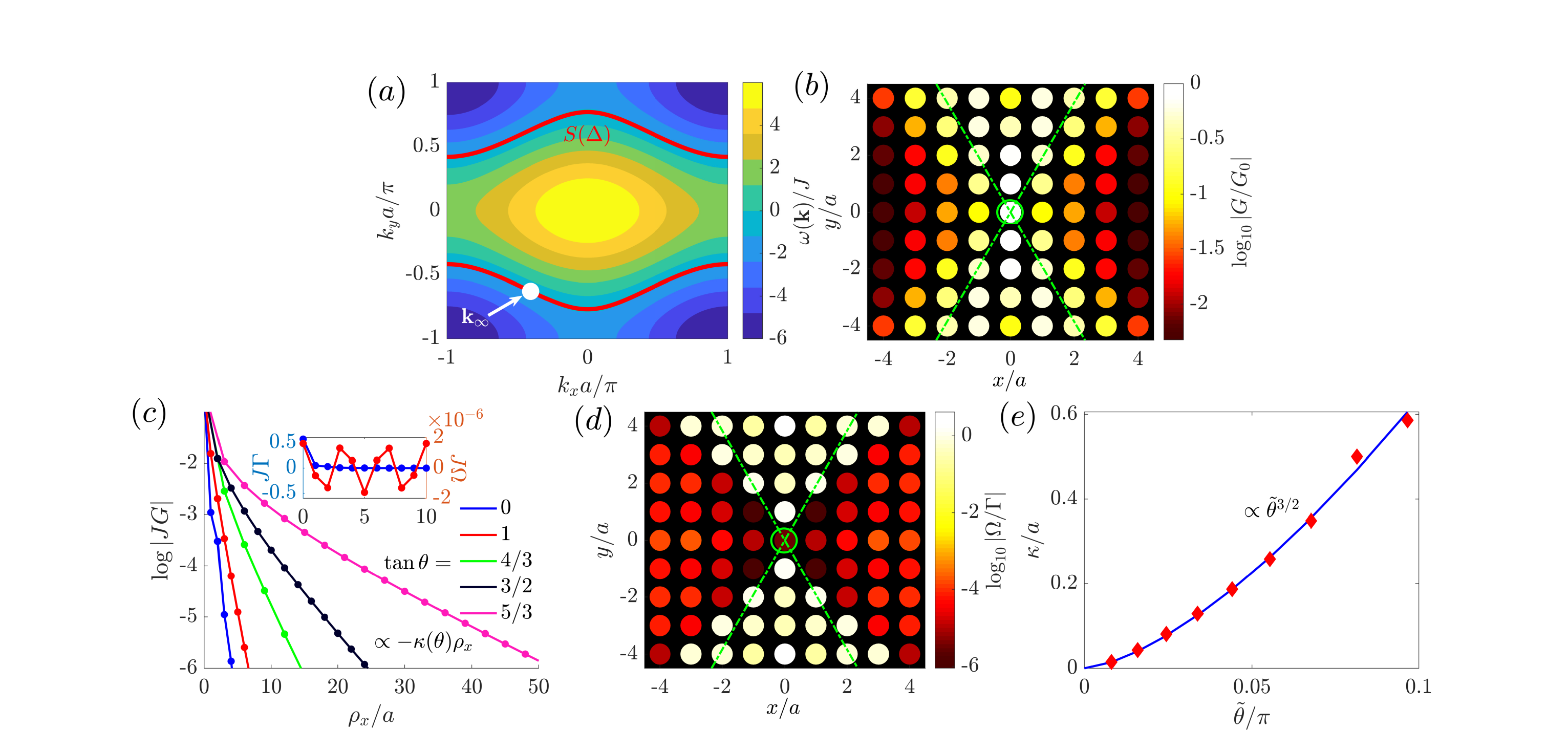}}
    \caption{(a) Dispersion relation for $J_{y} = 2J_{x} = 2J,$ with $\Delta = -J$. The white dot $\mathbf{k}_{\infty}$ denotes the line's inflection point with $\frac{\mathbf{v}(\mathbf{k}_{\infty}) \cdot \mathbf{m}(\mathbf{k}_{\infty}) \cdot \mathbf{v}(\mathbf{k}_{\infty})}{|\mathbf{m}(\mathbf{k}_{\infty})|} \to \infty$, where $\mathbf{v}(\mathbf{k}_{\infty})$ gives one of the caustic vectors (see (b)), upon which symmetry then gives the other. (b) Magnitude of the ratio of the Green's function (omitting the frequency argument) $G(\mathbf{0},\bm{\rho})$ to $G_{0} = G(\mathbf{0},\mathbf{0})$ on a log scale, where each dot denotes a lattice site. Dash-dotted green lines denote the caustics, given as the group velocity vector $\mathbf{v}(\mathbf{k}_{\infty})$ and its mirror image in the BZ. (c) $\log|JG|$ for varying $\tan\theta = \rho_{y}/\rho_{x}$. For reference, $\text{arg}[\mathbf{v}(\mathbf{k}_{\infty})] \approx 1.92$. Inset: $J\Omega, J\Gamma$ for $\theta = 0.$ (d) Ratio between $|\Omega|$ and $|\Gamma|$ on a log scale. (e) Dimensionless inverse decay length $\kappa/a$ of $
    \Gamma$ beyond the caustic. The diamonds (the solid line) give the value through exact calculation of the Green's function (the value obtained by fitting to \eqref{eq:ghost_decay_approx}).}\label{fig:anisotropy_square}
\end{figure*}

\section{The Green's function beyond caustics}

In our approximation $\eqref{eq:medrange},$ we see for larger $\rho$ that slow oscillations of the integrand arise with varying $\mathbf{k}$ when $\hat{\bm{\rho}}$ and $\hat{\mathbf{v}}$ become parallel. The integral in the vicinity of these stationary points then does not exponentially vanish with large $\rho$, i.e., the wavepackets with velocity parallel (including sign) to separation dominantly contribute to the integral, giving the usual stationary phase approximation (see Equation \eqref{eq:longrange}, \cite{stationary_optics,StatPhase2017,phonon_focus,photon_focus}). If the bath forbids wavepackets with particular group velocities, then we can expect the Green's function to vanish exponentially fast along these directions, which would invalidate~\eqref{eq:medrange} and require further analysis. With emitters weakly coupled to baths one typically observes emission into travelling waves for resonant frequency within the band~\cite{extended_dipole,cirac_noperturb2017,top_bath}, or evanescent waves when no travelling waves are available outside of the band~\cite{band_gap_1,band_gap_2,band_scaling,two_array_band}. However, when $d>1,$ anisotropy in the bath allows both regimes of decay to occur at a single frequency within the band due to the presence of wavepackets with group velocity in a particular direction and the absence in other directions, which results in the well-known `cone'-shaped emission into the bath (see Figure~\ref{fig:anisotropy_square}(b)). Beyond the caustics defining the cone boundary, the Green's function elements oscillate with a directionally dependent exponential decay. These so-called \textit{ghost waves} \cite{ghost_phase,ghost_wave} accompany hyperbolic dispersion \cite{hyperbolic,hyperbolic_disp,hyperbolic_disp_discrete}, and are distinct from the exponential decay (together with the partial bound states) of emitters with resonant frequency outside of the band. These waves are of interest in sub-wavelength imaging and low-loss sensing and transfer~\cite{ghost_polariton,ghost_wave}. When dispersion is hyperbolic-like (Figure~\ref{fig:anisotropy_square} (a)), our formula~\eqref{eq:medrange} is invalidated beyond the caustic. Indeed, in our approximation~\eqref{eq:medrange}, as $\hat{\bm{\rho}} \cdot \hat{\mathbf{v}}$ takes a constant sign for hyperbolic dispersion and $\bm{\rho}$ beyond a caustic, the amplitude $\Pi_{\rho}$ tends exponentially to 1 on all of $S$, so that the rapidly oscillating Bloch functions result in an exponential decay of the surface integral, i.e., the entire contribution to the Green's function is due to evanescent waves. This is also suggested by the stationary phase approximation of~\eqref{eq:medrange}, giving the well-known formula:
\begin{align}
	\label{eq:longrange}
	    |{G}| \sim 
	    \mathcal{A}\sum\limits_{\mathbf{k}_{0}}&\sqrt{\frac{|\mathbf{v}(\mathbf{k}_{0}) \cdot \mathbf{m}(\mathbf{k}_{0}) \cdot \mathbf{v}(\mathbf{k}_{0})|}{|\mathbf{m}(\mathbf{k}_{0})|}}\times\nonumber\\
	    &|\psi(\mathbf{k}_{0},\mathbf{r})\psi^{*}(\mathbf{k}_{0},\mathbf{r'})|\left(\frac{|v(\mathbf{k}_{0})|}{2\pi \rho}\right)^{\frac{(d-1)}{2}},
\end{align}
where the sum is taken over all $\mathbf{k}_{0}$ with $\mathbf{v}(\mathbf{k}_{0})$ parallel to $\bm{\rho}$ (including sign). Here we have the effective mass tensor $\mathbf{m}$ with  $[\mathbf{m}^{-1}(\mathbf{k})]_{ij} = \frac{\partial^2 \omega}{\partial k_{i}\partial k_{j}}$ and $|\mathbf{m}(\mathbf{k}_{0})|$ denoting its determinant. We note (see Appendix~\ref{app:long-range:stationary-phase}) that the term within the square root corresponds to the inverse of the Gaussian curvature of $S(\Delta)$ at $\mathbf{k}_{0}$, giving the more commonly known form~\cite{stationary_optics,StatPhase2017,phonon_focus,photon_focus}. When $\bm{\rho}$ lies beyond a caustic (Figure~\ref{fig:anisotropy_square} (b)) the sum in~\eqref{eq:longrange} is empty and the naive approximation predicts a zero Green's function. However, heading beyond the caustic allows us to the access exponential decay of oscillating evanescent waves whose origin is fundamentally distinct from the exponential localization of bound states observed for emitters resonant outside a band. Adjusting $\bm{\rho}$ beyond the caustic modulates the exponential decay between emitters akin to dynamics near a band edge~\cite{hood_green,two_array_band,band_scaling}. Specifically, in Figure \ref{fig:anisotropy_square}(c), the exponential decay length of $G$ can be tuned between zero (at the caustic) and a maximum value determined by bath microscopics. Whilst $\Omega$ must now be calculated exactly using the contributions from the entire BZ, a universal scaling behaviour near the caustic can generally be obtained for $\Gamma$. Assuming the angle $\tilde{\theta}$ between $\bm{\rho}$ and the caustic is small with $\bm{\rho}$ lying just \textit{outside} the light cone (Figure~\ref{fig:anisotropy_square} (b)), a saddle point analysis detailed in Appendix~\ref{app:critical-angle} gives a generic result in 2D 
\begin{equation}\label{eq:ghost_decay_approx}
\Gamma \sim \frac{e^{-\kappa \rho}}{\rho^{1/2}}, \ \ \ \ \ \ \kappa \propto a|\tilde{\theta}|^{3/2},
\end{equation}
as observed in Figures \ref{fig:anisotropy_square}(c-e). The resulting exponential decay is consistent with the heuristic result obtainable by choosing a complex $
\mathbf{k}$ (corresponding to a complex pole) in \eqref{eq:longrange}, recalling the complex exponential of $\mathbf{k}$ implicit in $\psi(\mathbf{k}_{0},\mathbf{r})$. In the square lattice we typically find $|\Omega|  \ll |\Gamma|$ (Figure~\ref{fig:anisotropy_square}(d)) for almost all lattice sites outside of the light cone. This is in contrast with bound states for emitters resonant outside the band, where $0 = |\Gamma| \ll |\Omega|.$ Because $|Omega|$ is far smaller in magnitude than $\Gamma,$ we have $|G|\sim |\Gamma|$ and thus find that $\Gamma$ shares the same scaling as $|G|,$ whose magnitude is shown in Figure \ref{fig:anisotropy_square}(c). In particular we plot in the inset the result for $\bm{\rho} \propto \mathbf{x}.$ Here $\Omega$ is many orders of magnitude smaller than $\Gamma$, which itself is rapidly exponentially decaying. This feature allows one to construct nearest-neighbor dissipative spin models with purely incoherent interactions. Furthermore, we can see that $\Omega$ is dominated by $\Gamma$ almost everywhere beyond the caustic (Figure~\ref{fig:anisotropy_square}(d)), so that one has a large freedom over the decay length in the dissipative models whilst keeping $\Omega$ relatively suppressed. Equation~\eqref{eq:ghost_decay_approx} and the results of Figure~\ref{fig:anisotropy_square} (b-e) are the second key results of this work. In emitters coupling to structured media the detuning and the angle can be freely varied \cite{band_scaling,two_array_band}, giving one dynamical control over the decay length of interactions beyond caustics similar to the $\Delta^{-1/2}$ scaling of decay length near a band edge~\cite{band_scaling,two_array_band,hood_green,directional_gap}. We finally note that in a scalar bath the results we have obtained cannot be predicted by the continuum medium approximation~\cite{hyperbolic_disp_discrete}, which results in $\Gamma = 0$ beyond the caustic. The discreteness of the lattice is thus the key to observing oscillating evanescent waves with with $|\Gamma| \gg |\Omega|.$ 

\section{Emission into magnetic hyperbolic baths}
\subsection{Quasi-1D transport}

We now turn our attention to 2D, although one can also generalize to higher dimensions. The final case we consider beyond the previous two sections is that of a hyperbolic bath with quasi-broken periodicity due to the effect of an artificial magnetic field acting on the lattice. We here exactly simulate emission into the bath as the Green's function may become arbitrarily degenerate as the effective magnetic field strength is varied.
Away from Van Hove singularities, we can characterise $S(\Delta)$ as an \textit{open} or \textit{closed} orbit, which refers to whether wavepackets resonant at $\Delta$ traverse an open trajectory or closed loop in the presence of an (artificial) magnetic field~\cite{open_orbit}. Figure~\ref{fig:initialcompare}(b) gives a closed orbit. However, in the case of Figure \ref{fig:anisotropy_square}(a), one observes open orbits corresponding to the upper and lower branches of $S(\Delta)$. We probe the effect of these orbits on resonant emission, using an effective magnetic field on the lattice. Assuming an absence of Berry curvature, valid for our square lattice, the semi-classical equations of motion for a wavepacket are \cite{semiclassical_electron,semiclassical_photon}
\begin{eqnarray}\label{eq:equations_of_motion}
    &\dot{\mathbf{r}} = \mathbf{v}(\mathbf{k}), \\
    &\label{eq:k_motion}\dot{\mathbf{k}} = -\mathbf{v}(\mathbf{k}) \times {\mathbf{B}}.
\end{eqnarray}
We take the effective magnetic field $\mathbf{B} = \frac{\Phi}{a^2}\hat{\mathbf{z}} = B\hat{\mathbf{z}}$ pointing out of the lattice, corresponding to the physical effect of a gauge field $\mathbf{A}(\mathbf{r}) = \frac{\mathbf{r} \times \hat{\mathbf{z}}}{2a^2}$ resulting in phase accumulation $J_{\mathbf{r}_{i}\mathbf{r}_{j}} \to J_{\mathbf{r}_{i}\mathbf{r}_{j}}e^{i\phi_{ij}}$, with $\phi_{ij} = \int_{\mathbf{r}_{i}}^{\mathbf{r}_{j}}\mathbf{A}(\mathbf{r})\cdot d\mathbf{r}$ and $\Phi = \sum_{\square} \phi_{ij}$ for a plaquette $\square$. Conservation of $\omega(\mathbf{k})$ through Equation \eqref{eq:k_motion} dictates that the wave packet traces  in $\mathbf{k}$ space along $S$, whose geometry will be imprinted into evolution in $\mathbf{r}$ space. While circular Landau orbits are obtained for closed orbits of isotropic regimes~\cite{photon_polariton,landau_circle}, open orbits instead result in a finite transport of wave packets over a period. The spatial (temporal) period $\mathbf{l} \ (\tau)$ of orbit trajectories is characterised as
\begin{eqnarray}
    &\label{eq:magnetic_spatial_period}\mathbf{l} = \int d\mathbf{r} = \frac{1}{B}\int_{S}\frac{d\mathbf{k}\mathbf{v}(\mathbf{k})}{v(\mathbf{k})} = \frac{\hat{\mathbf{a}}_{y}}{\alpha}, \\
    &\label{eq:magnetic_temporal_period}\tau = \int dt = \frac{1}{B}\int_{S}\frac{d\mathbf{k}}{v(\mathbf{k})} = \frac{\Gamma }{2\alpha},
\end{eqnarray}
where $\alpha  = \frac{\Phi}{2\pi}$~\cite{photon_polariton}. $\mathbf{l}$ is invariant under microscopic variations of the lattice (for fixed $\hat{\mathbf{a}}_{y}$ and field) until a Van Hove singularity is encountered, so that such quasi-1D orbits are expected to be present generally when band transitions and Berry curvature can be neglected~\cite{semiclassical_electron}.
\begin{figure}[t!]
    \includegraphics[scale=0.22]{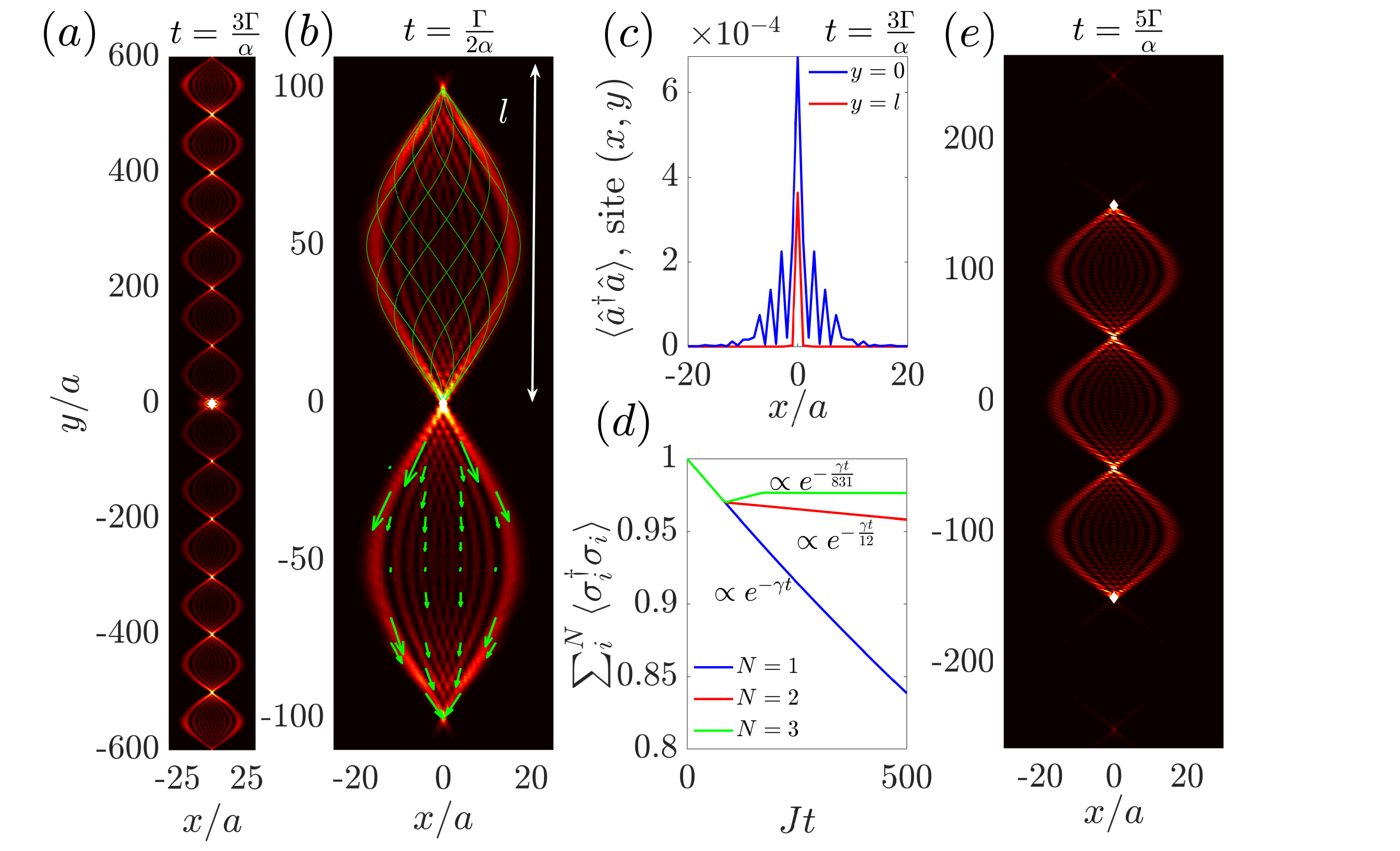}
    \caption{Emission into the magnetic bath 111 (1201) sites across in the $x (y)$ direction. Parameters are as in Figure \ref{fig:anisotropy_square}, with $\alpha = 0.01.$ In (a-c), $g=0.1J,$ and in (d-e), $g=0.025J.$ (a) Bath population at $t = 6\tau$ for an emitter prepared in the excited state, and located at the central white diamond. (b) Bath population at $t=\tau$. Green lines give semiclassical trajectories evolving from the emitter with $\mathbf{k}$ sampled from $S$. Arrows correspond to photonic current~\cite{photon_polariton}. (c) Bath population along the ${x}$ axis when ${y} = 0, l.$ (d) Total emitter population for one, two, and three emitters, each separated by $3l.$ The two-atom (three atom) subradiant state is prepared. (e) Bath population at $t=10\tau$ for two emitters separated a distance $3l$ and prepared as in (d). }\label{fig:magnetic_orbits}
\end{figure} 
Phenomena due to non-zero $l = |\mathbf{l}|$ are manifest when an emitter is coupled to the bath, as emission is dominated by wavepackets localized on $S$ (as in Equation \eqref{eq:medrange}) that evolve as Equations (10-13). In Figure \ref{fig:magnetic_orbits}, the population plots (a-b) confirm quasi-1D transport and are consist with the semiclassical orbits. These findings summarized in Figure~\ref{fig:magnetic_orbits} are the third key results of this work. Notably, periodicity of the travelling orbits combined with common initial position at the emitter demands periodic refocussing of emission. In Figure \ref{fig:magnetic_orbits}(c), refocussing of emission at subsequent sites occurs almost down to the single site level, which results in strong coupling of emitters placed with separations of integer multiples of the spatial period. Exploiting this single site resolution and separating emitters by integer multiples of $l$ in the $y$ direction can be used to form an effective 1-d cavity in the 2-d bath when emitters are prepared in a dark state~\cite{sinha_nonmarkov,chang_cavity}.
A snapshot of the bath population at $t=10\tau$ for the two emitters in Figure \ref{fig:magnetic_orbits}(e) displays destructive interference forming bound states in the continuum~\cite{chang_cavity,sinha_nonmarkov,bound_continuum}. Significant emitter and localized bath population between emitters can be sustained orders of magnitude beyond the single-emitter lifetime, where fidelity increases with emitter number as shown in Figure \ref{fig:magnetic_orbits}(d). However, the cavities modes inherently comprise a continuum of $\mathbf{k}$ modes on $S$. Additionally, population nodes do not occur at the edges of the cavity, in contrast to the recently established correspondence between coupled emitters and site vacancies~\cite{not_vacany}, although the details of these effective cavity modes will be investigated in a separate work. We finally note that the discreteness of the lattice (i.e., the finite size of the Brillouin zone) is also crucial for observing these effects, as the periodic `reflection' of wave packets that confines them to one dimension occurs due to the presence of inflection points ($\mathbf{k}_{\infty}$ in Figure~\ref{fig:anisotropy_square}(a)) on $S(\Delta)$, which are absent in the case of continuum hyperbolic dispersion. 

\subsection{Topological robustness}

\begin{figure}[t!]
    \centering
    \includegraphics[scale=0.24]{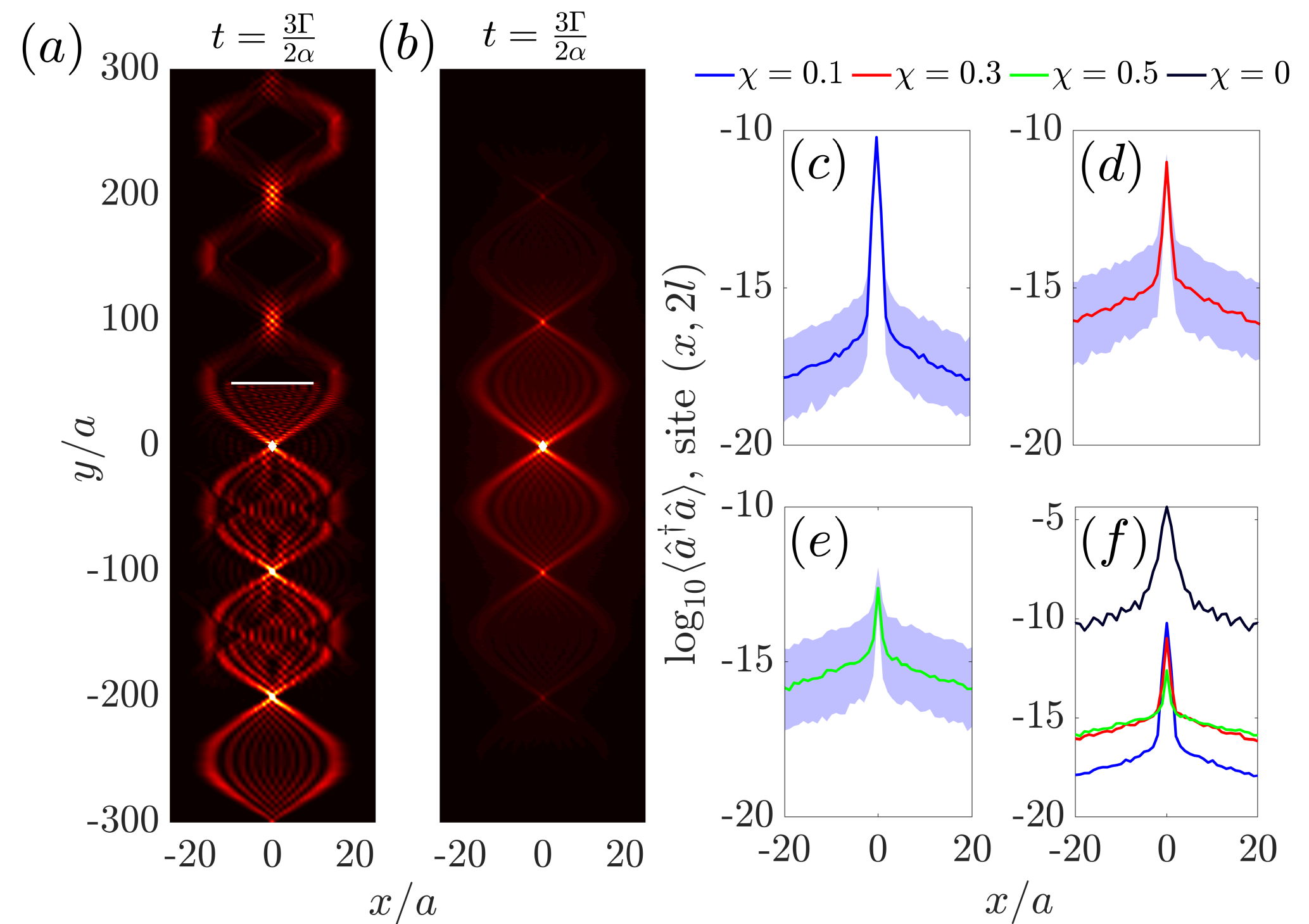}
    \caption{Parameters are as in Figure \ref{fig:magnetic_orbits} with $g=0.025J$. (a) Bath population at $t = 3\tau$ in the presence of an obstruction (white line) at $y=l/2, x \in [-10a,10a]$ obtained by decoupling all obstruction sites from the lattice. (b) Averaged bath population at time $t = 3\tau$ for disorder $\chi = 0.5.$ (c-f) Average log-populations for the cross-section $y = 2l$ in the presence of varying disorder parameter $\chi.$ The solid lines give the average whilst the shadow denotes a standard deviation. In (f) the population in the presence of no disorder corresponding to Figure \ref{fig:magnetic_orbits}(a) is shown for comparison.}
    \label{fig:topological_robustness} 
\end{figure}

We finally investigate the robustness of the quasi-1D transport in the system. Waves elastically scattered from local perturbations are also restricted to quasi-1D, and in Figure \ref{fig:topological_robustness}(a) we observe bidirectional scattering, together with periodic refocussing of the beam after impinging on an obstruction, similar to the `healing' observed in Bessel and Airy beams~\cite{bessel_beam,airy_healing}, albeit effected by a different mechanism. The finding of Figure~\ref{fig:topological_robustness} are the final key results of this work. In addition to local obstructions, quasi-1D propagation is also robust to global lattice perturbations. Introducing on-site energy disorder $XJ,$ where $X$ is a random variable sampled uniformly in $[-\chi, \chi],$ for the disorder parameter $\chi$ and averaging over 500 realizations, the average bath populations are given in Figure \ref{fig:topological_robustness}(b-f). In (c-f) we find the periodic refocussing is qualitatively preserved, with bath population maintaining its peak by many orders of magnitude at the central site. In (b) the profile averaged over all realisations is additionally sustained. Fluctuations in the weakly populated sites are additionally many orders of magnitude smaller than the central site population, which itself sees small variations up to moderate disorder $\chi = 0.5.$ Notably, the exponential decay rate away from the central site is additionally shared across all disorders as in (f). These effects are a consequence of the topological robustness of open orbits protecting quasi-1D emission in an application of the Fermi surface topology \cite{geometry_van_hove,fermi_surface_1,fermi_surface_2} of condensed matter physics.
\\
\section{Conclusion}
To summarize, we have studied the emission into structured media, exploring the consequences of (quasi-)breaking of periodicity and anisotropy in the effective bath. We have proven a decomposition of linear wave propagation in arbitrary periodic local lattices as a sum of a travelling part and an evanescent part. The approximation can fail beyond caustics when hyperbolic-like dispersion is present, where oscillating evanescent waves are the dominant transport mechanism. Here we found a universal long-range scaling of the Green's function near caustics and revealed that the Green's function can be dominantly dissipative in a large region beyond caustics, allowing the creation of purely dissipative spin-models with interaction decay length tunable via the angle at which emitters are placed. Finally, breaking periodicity with an effective magnetic field resulted in topologically protected quasi-1D transport in direct contrast to the usual Landau orbits, offering robust bidirectional transport and atypical bound states in the continuum. Of particular interest for continuations would be more detailed studies investigating transport around Van Hove singularities and degeneracies~\cite{tunable_tudela,tudela_dirac,cirac_noperturb2017}, whilst a study of dynamics in conjunction with standard topological phenomena could prove particularly fruitful~\cite{semiclassical_electron,tunable_tudela,top_bath,not_vacany}. Whether more general statements regarding the decay of the Green's function beyond caustics can be made also remains an open question.

\section{Acknowledgements}

\begin{acknowledgements}This work was supported by the Okinawa Institute of Science and Technology. Simulations in this work were performed using the open source Julia packages DifferentialEquations.jl~\cite{diff_jl}, QuadGK.jl~\cite{quadgk_jl}, and MDBM.jl~\cite{MDBM_jl}. The authors acknowledge discussions with F. le Kien. We are additionally grateful for the scientific cluster and resources provided by the Scientific Computing and Data Analysis section of the Research Support Division at OIST.
\end{acknowledgements}

\appendix

\section{\label{app:long-range}Derivation of the long-range behaviour of the Green's function}

\subsection{\label{app:long-range:nodegen}Proving that degeneracies contribute only exponential decay}

When site couplings are local (i.e.~a site only couples to finitely many neighbours), the dispersion relation is analytic on the real axis \cite{bloch_analytic} away from degeneracies in the general case of multiple bands, which we index by $\nu$. Away from Van Hove singularities $\nabla \omega_{\nu}(\mathbf{k}) \neq 0$, the implicit function theorem applies, and the resonant set $S_{\nu}$ is a smooth manifold of co-dimension $(d-1),$ allowing one to apply the full power of differential geometry. However, at a degeneracy, it seems that the integrand of 
\begin{equation}
    I = \sum_{\nu}\int_{BZ}\frac{d^{d}\mathbf{k}}{(2\pi)^d}\frac{\psi_{\nu}(\mathbf{k},\mathbf{r})\psi^{*}_{\nu}(\mathbf{k},\mathbf{r}')}{\Delta - \omega_{\nu}(\mathbf{k})}
\end{equation}
becomes non-smooth \cite{katoperturbation2013}, contributing a power-law decay. Assuming the degeneracy does not coincide with the resonant set $S$, we show that summing over all bands removes any non-analyticity present in individual bands. This allows us to consider only a single band in the main text when the resonant set is away from degeneracies. The derivation follows in a manner similar to \cite{sanchez_limits}. Recall the unitary transformation between eigenfunctions $\psi_{i}$ and the Bloch wave corresponding to an individual sublattice:
\begin{equation}
    |\psi_{\nu}(\mathbf{k})\rangle = \sum_{j}U_{\nu j}(\mathbf{k})|\mathbf{k}_{j}\rangle,
\end{equation}
where $|\mathbf{k}\rangle_{i} = \frac{1}{\sqrt{N}}\sum_{r_{i}}e^{i\mathbf{k}\cdot \mathbf{r}_{i}}|\mathbf{r}_{i}\rangle$ with $|\mathbf{r}_{i}\rangle$ the localized excitation at any lattice site $\mathbf{r}_{i}$ in lattice $i$. In the thermodynamic limit we conversely have $|\mathbf{r}_{i}\rangle = \int_{BZ}\frac{d^{d}\mathbf{k}}{(2\pi)^d}e^{-i\mathbf{k}\cdot \mathbf{r}_{i}}|\mathbf{k}_{i}\rangle,$ and may rewrite the inner product wavefunctions:
\begin{eqnarray}
    &\psi_{\nu}(\mathbf{k},\mathbf{r}_{i}) = \langle \mathbf{r}_{i} | \psi_{\nu}(\mathbf{k}) \rangle = \nonumber\\ &\sum_{j}\int_{BZ}\frac{d^{d}\mathbf{k}'}{(2\pi)^d}e^{i\mathbf{k}\cdot \mathbf{r}_{i}}U_{\nu j}(\mathbf{k})\langle \mathbf{k}'_{i}|\mathbf{k}_{j}\rangle = e^{i\mathbf{k}\cdot \mathbf{r}_{i}}U_{\nu i}(\mathbf{k}).
\end{eqnarray}
    Assuming $\mathbf{r}$ and $\mathbf{r}'$ belong to lattices $i$ and $i'$ respectively, we rewrite the integrand
\begin{equation}
    I = \sum_{\nu}\int_{BZ}\frac{d^{d}\mathbf{k}}{(2\pi)^d}\frac{e^{i\mathbf{k}\cdot \bm{\rho}}U_{i' \nu}^{*}(\mathbf{k})U_{\nu i}(\mathbf{k})}{\Delta - \omega_{\nu}(\mathbf{k})},
\end{equation}
or more succinctly in matrix notation
\begin{equation}
    I = \int_{BZ}\frac{d^{d}\mathbf{k}}{(2\pi)^d} e^{i\mathbf{k} \cdot \bm{\rho}}\big[\mathbf{U}^{\dagger}(\mathbf{k})(\Delta - \mathbf{D}(\mathbf{k}))^{-1}\mathbf{U}(\mathbf{k})\big]_{i' i}.
\end{equation}
We note the diagonalization $\mathbf{h}(\mathbf{k}) = \mathbf{U}^{\dagger}(\mathbf{k})\mathbf{D}(\mathbf{k})\mathbf{U}(\mathbf{k})$ for diagonal $[\mathbf{D}(\mathbf{k})]_{\nu \nu} = (\Delta - \omega_{\nu}(\mathbf{k}))^{-1}$ and recognize the inner element as an inverse of the finite dimensional resolvant 
\begin{equation}
       I = \int_{BZ}\frac{d^{d}\mathbf{k}}{(2\pi)^d} e^{i\mathbf{k} \cdot \bm{\rho}}(\Delta - \mathbf{h}(\mathbf{k}))^{-1}_{i' i}.
\end{equation}
where $\mathbf{h}(\mathbf{k})$ is now the finite dimensional matrix with analytic elements describing coupling between each of the Bloch sublattices. By standard resolvant analysis \cite{katoperturbation2013}, the resolvant is analytic in both, $\mathbf{k}$ and $\Delta$ whenever $\Delta$
 is not equal to any of the eigenvalues of $\mathbf{h}(\mathbf{k}),$ i.e., not equal to any of the energy bands at $\mathbf{k}.$ Thus, when the resonant level set energy $\Delta$ does not lie at a degeneracy, the integrand is analytic with respect to integration variable $\mathbf{k}$ away from any resonant level sets, and localization of the full Green's function integral around a degeneracy but away from the resonant level set produces an analytic integrand allowing for asymptotic analysis purely around the resonant level set as in the following sections.

 \subsection{\label{app:long-range:coherent-part} Long range behaviour of the coherent part of the Green function in $d$ dimensions}
Consider the Green's function describing propagation through a single mode $\nu$ (whose subscript we omit in the following) of a periodic continuum bath:
\begin{equation}
\label{appeq:greenfunction}
	{G}(\mathbf{r},\mathbf{r}',\Delta) = \mathcal{A} \int_{\text{BZ}}\frac{d^{d}\mathbf{k}}{(2\pi )^d}\frac{\psi(\mathbf{k},\mathbf{r})\psi^{*}(\mathbf{k},\mathbf{r'})}{\Delta - \omega(\mathbf{k}) + i0},
\end{equation}
The integral is understood through the limiting absorption principle, taking a small imaginary part in the denominator to zero. Whilst the wave function $\psi$ is in this case a scalar, we note that similar expressions arise when $\psi$ is a tensor, i.e.~when polarization is available in an electromagnetic system. With time reversal symmetry of the bath assumed, we extract the strongly singular continuum integral corresponding to coherent wave propagation:
\begin{eqnarray}
\label{appeq:om}{\Omega} = \mathcal{A}\mathcal{P}\int_{\text{BZ}}\frac{d^{d}\mathbf{k}}{(2\pi )^d}\frac{\psi(\mathbf{k},\mathbf{r})\psi^{*}(\mathbf{k},\mathbf{r'})}{\Delta - \omega(\mathbf{k})},
\end{eqnarray}
and proceed with an asymptotic large $\bm{\rho} = \mathbf{r} - \mathbf{r}'$ treatment, where the direction vector of $\bm{\rho}$ is kept fixed. Having shown analyticity around degeneracies, we now address the  (near-)resonant contributions from $S$. Under the assumption of locality, we may take $S$ to be a smooth manifold of codimension one, and note that $\omega(\mathbf{k})$ is analytic for real $\mathbf{k}$ away from degeneracies. With this we may use a partition of unity (or localization argument). That is, we may choose smooth bump functions \cite{2015BumpFourier} with support on arbitrarily small regions near $S$ and away from $S$ respectively. Away from $S$ the integrand is smooth, and so exponential decay of the integrand is obtained. Let us first factorize the Bloch waves:
\begin{equation}
{\Omega} = \mathcal{A}\mathcal{P}\int_{\text{BZ}}\frac{d^{d}\mathbf{k}}{(2\pi )^d}\frac{\Phi(\mathbf{k})e^{i\bm{\rho}\cdot \mathbf{k}}}{\Delta - \omega(\mathbf{k})},
\end{equation}
where we increase the separation with $\mathbf{r}, \mathbf{r}'$ fixed modulo the lattice, so that $\Phi(\mathbf{k}) = \psi(\mathbf{k},\mathbf{r})\psi^{*}(\mathbf{k},\mathbf{r'})e^{-i\bm{\rho}\cdot \mathbf{k}},$ $\Phi$ is unchanging with $\bm{\rho}$ and can be considered purely as a function of $\mathbf{k}.$ We may parametrise the vicinity of the resonant set $S = S(\Delta) = \{\mathbf{k} : \omega(\mathbf{k}) = \Delta\}$ using canonical (Fermi) coordinates in a \textit{tubular neighbourhood} of $S$:
\begin{equation}
	\mathbf{k} = (t,\bm{\xi}) = \bm{\xi} + t\hat{\mathbf{v}}({\bm{\xi}}),
\end{equation}
where  $\bm{\xi}$ parameterizes the level  set $S$, and $\hat{\mathbf{v}}$ is the unit normal, which we assume faces outwards. As the level set is a smooth surface of codimension $1$, it is orientable and such a normal always exists. Note that as the group velocity is assumed to be non-vanishing on the surface, the velocity vector always points inside or outside. We assume without loss of generality that the group velocity points outside, although a similar result is obtained otherwise. The above parametrisation is smooth when $t$ is smaller than any of the radii of curvature anywhere on $S$ \cite{WeylVolume2017}, and we choose $\epsilon$ such that $|t| < \epsilon$ satisfies this condition.

We now use the partition of unity to isolate the contribution from $S$. For this we consider the integral:
\begin{equation}
R = \mathcal{A}\mathcal{P}\int_{\delta S} \frac{d^d \mathbf{k}}{(2\pi)^d} \phi_{\delta S}(\mathbf{k})\frac{\Phi(\mathbf{k})e^{i\bm{\rho}\cdot \mathbf{k}}}{\Delta - \omega(\mathbf{k})},
\end{equation}
where $\delta S$ is a thin tubular region surrounding $S$, with $|t| < \epsilon$. This is smoothly facilitated by the bump function $\phi_{\delta S} \in C^{\infty}$ forming one of our partition of unity. The bump function can be given entirely through dependence on $t$:
\begin{equation}
    \phi_{\delta S}(\mathbf{k}) = \phi_{\epsilon}(t),
\end{equation}

where $\phi_{\epsilon}(t)$ is the usual one-dimensional bump function, normalized to unity at zero:
\begin{equation}
    \phi_{\epsilon}(t) = \begin{cases} 
      \exp(\frac{t^2}{t^2-\epsilon^2}) & |t|  < \epsilon \\
      0 & \text{Otherwise}
   \end{cases}
\end{equation}

We now change coordinates systems, giving
the volume element obtained in the derivation of\textit{Weyl's tube formula} \cite{WeylVolume2017,gray_tubes,tube_backup,tube_backup_two}:
\begin{equation}
	d^{d}\mathbf{k} = \det (\mathbf{I} + t \mathbf{K})d\bm{\xi}dt,
\end{equation}

where $\mathbf{K}$ is the \textit{second fundamental form} of the surface $S$. In the case of an implicit surface we have \cite{secondform}
\begin{equation}
	\mathbf{K} = \mathbf{K}(\bm{\xi}) = -\frac{\mathbf{H}_{T}(\bm{\xi})}{v(\bm{\xi})},
\end{equation}

for the Hessian 
\begin{equation}
	[\mathbf{H}(\mathbf{k})]_{ij} = \frac{\partial^{2}\omega(\mathbf{k})}{\partial k_{i} \partial k_{j}},
\end{equation}

which we restrict to a linear map $\mathbf{H}_{T}(\bm{\xi})$ acting on the tangent space of $S$ at $\bm{\xi}.$ We obtain
\begin{equation}
	R = \mathcal{A}\mathcal{P}\int_{S}\frac{e^{i\bm{\xi}\cdot\bm{\rho}}d^{d-1} \bm{\xi}}{(2\pi)^{d-1}}\int_{-\epsilon}^{\epsilon} \frac{dt}{2\pi}\phi_{\epsilon}(t) \frac{\Phi(t,\bm{\xi})e^{it\bm{\rho}\cdot \hat{{\mathbf{v}}}}}{\Delta - \omega(t,\bm{\xi})}\det (\mathbf{I} + t \mathbf{K}).
\end{equation}

We note that the only non-smoothness is at $t=0$ and  rewrite
\begin{eqnarray}
	&R =\nonumber\\
    &\mathcal{A}\int_{S}\frac{e^{i\bm{\xi}\cdot\bm{\rho}}d^{d-1} \bm{\xi}}{(2\pi)^{d-1}}\mathcal{P}\int_{-\epsilon}^{\epsilon} \frac{dt}{2\pi} \frac{L(t,\bm{\xi})e^{it\bm{\rho}\cdot \hat{{\mathbf{v}}}}\phi_{\epsilon}(t)}{t}\det (\mathbf{I} + t \mathbf{K})\nonumber\\
	&
\end{eqnarray}

Here, $L(t,\bm{\xi}) = \frac{t\Phi(t,\bm{\xi})}{\Delta - \omega(t,\bm{\xi})}$ defines a smooth function within $\delta S$ (local expansion in small $t$ reveals that smoothness can be extended to $t=0$) as group velocity is assuming non vanishing, so that the denominator only scales as $1/t$. Additionally, the determinant may be expanded, upon which we need only the lowest order term (the higher order terms cancel $t$s and produce exponential decay). We can separate out only the contribution at $t=0$, and discard small contributions to obtain

\begin{equation}
\label{appeq:discarded}
	\Omega \sim R \sim  \mathcal{A}\int_{S}\frac{e^{i\bm{\xi}\cdot\bm{\rho}}d^{d-1} \bm{\xi}}{(2\pi)^{d-1}}\mathcal{P}\int_{-\epsilon}^{\epsilon} \frac{dt}{2\pi} \frac{L(0,\bm{\xi})e^{it\bm{\rho}\cdot \hat{{\mathbf{v}}}}\phi_{\epsilon}(t)}{t},
\end{equation}
plus corrections of the order $O(\rho^{-\infty})$ as $\rho \to \infty$, and $L(0,\bm{\xi})$ is understood as a limit. We use $O(\rho^{-\infty})$ to denote decay faster than any polynomial (superpolynomial) which in the main text we see corresponds to exponential decay. The remainder we subtracted formed a smooth function in $\delta S$, and for $|t|<\epsilon/2$  the surface element is also smooth, so that the remainder will decay superpolynomially by the principal of non-stationary phase~\cite{StatPhase2017}. Thus, replacing $L(0,\bm{\xi}) = -\frac{\Phi(0,\bm{\xi})}{v(\bm{\xi})}$ we may return to the standard coordinate system and collect together the $t$ dependent variables to write
\begin{eqnarray}
  	&\Omega = -\frac{i\mathcal{A}}{2}\int_{S}\Theta_{\rho}(\hat{\bm{\rho}}\cdot\hat{\mathbf{v}}(\mathbf{k}))\frac{\psi(\mathbf{k},\mathbf{r})\psi^{*}(\mathbf{k},\mathbf{r'})d^{d-1} \mathbf{k}}{v(\mathbf{k})(2\pi)^{d-1}} \nonumber\\
  	&+ O(\rho^{-\infty}),
\end{eqnarray}

where we have relabelled $\bm{\xi} \in S \to \mathbf{k} \in S$ and where we have the phase function after discarding the vanishing principal value integral involving cosine and a change of variables:
\begin{equation}
\label{appeq:phasefunc}
	\Theta_{\rho }(x) = -i\mathcal{P}\int_{-\epsilon}^{\epsilon}\frac{e^{i\rho xy}\phi_{\epsilon}(y)}{\pi y} =  \int_{-\infty}^{\infty} \frac{dy \sin(xy)\phi_{\rho \epsilon}(y)}{\pi y}.
\end{equation}
of the main text. As the integral of the Fourier transform of the bump function \cite{2015BumpFourier}, has the asymptotic behaviour
\begin{eqnarray}
  &\Theta_{\rho}(x) = \text{Sign}(x) +  O\big(\text{Erf}(|\rho\epsilon x|^{3/4})\big) \nonumber\\
  &= \text{Sign}(x) + O\big(|\rho\epsilon x|^{-1/4}\exp(-|\rho\epsilon x|^{1/2})\big)
\end{eqnarray}
with $\epsilon > 0$ and $x \to \pm \infty.$ We thus have the full Green's function approximant of the main text, converging to $G$ with exponentially small errors as $\bm{\rho} \to \infty:$
\begin{equation}
	\label{appeq:bestapprox}
	G \sim -\frac{i\mathcal{A}}{2}\int_{S}(1+\Theta_{\rho}(
	\hat{\bm{\rho}}\cdot\hat{\mathbf{n}}))\frac{\psi(\mathbf{k},\mathbf{r})\psi^{*}(\mathbf{k},\mathbf{r'})d^{d-1} \mathbf{k}}{v(\mathbf{k})(2\pi)^{d-1}}.
\end{equation}

\subsection{\label{app:long-range:stationary-phase}Stationary phase approximation on the resonant level set}

Here we connect our derivation to the standard stationary phase approximation. To evaluate the asymptotics 
\begin{equation}
	\label{app:bestapprox}
	G \sim -\frac{i\mathcal{A}}{2}\int_{S}(1+\Theta_{\rho}(
	\hat{\bm{\rho}}\cdot\hat{\mathbf{n}}))\frac{\psi(\mathbf{k},\mathbf{r})\psi^{*}(\mathbf{k},\mathbf{r'})d^{d-1} \mathbf{k}}{v(\mathbf{k})(2\pi)^{d-1}},
\end{equation}

in the geometric optics limit, we make use of the stationary phase argument. Whilst the phase function $\Theta_{\rho}$ is also rapidly oscillating with $\rho,$ integration by parts shows that for our choice of $\epsilon$ small enough, the oscillations due to the exponent are dominant, and the leading contribution to the integral come from points where the exponential argument is at an extremum on $S$. The integral rapidly converges to a typical exponential integral, and we use the general stationary phase formula on a manifold~\cite{StatPhase2017}:
\begin{eqnarray}
 \label{appeq:manifoldstatphase}
    &\int_{S}d^{d-1}\mathbf{k}f(\mathbf{k})e^{i\lambda \theta(\mathbf{k})} \nonumber\\
    &\sim \Big(\frac{2\pi}{\lambda}\Big)^{(d-1)/2}\sum_{\mathbf{k}_{0}}f(\mathbf{k}_{0}) \frac{e^{\text{Sgn}(H(\mathbf{k}_{0}))\frac{\pi i}{4}}e^{i\lambda \theta(\mathbf{k}_{0})}}{\sqrt{|\det[H(\mathbf{k}_{0})]|}},
\end{eqnarray}
plus terms of order $O(\lambda^{(d-3)/2})$ as $\lambda \to \infty$, where the sum is over (assumed) discrete critical points $\mathbf{k}_{0}$ of $\theta$ on $S.$ The Hessian $H$ is also computed as a Hessian on the manifold $S.$ When applied to \eqref{appeq:bestapprox}, we have simply the phase function $\theta(\mathbf{k}) = \mathbf{k} \cdot \hat{\bm{\rho}},$ the height function with respect to $\hat{\bm{\rho}}.$ According to \cite{height2016}, the Hessian of the height function restricted to $S$ and at a critical point is simply given as the second fundamental form, $H(\mathbf{k}_{0}) = -\mathbf{K}(\mathbf{k}_{0})$, so that the denominator is simply the square root of $|K(\mathbf{k}_{0})|,$ for the Gaussian curvature $K$. Plugging in \eqref{appeq:bestapprox} gives the final long-range approximation:
\begin{equation}
    G \sim -i\mathcal{A}\frac{\psi(\mathbf{k}_{0},\mathbf{r})\psi^{*}(\mathbf{k}_{0},\mathbf{r'})e^{\frac{i\pi\text{Sgn}[K(\mathbf{k}_{0})]}{4}}}{v(\mathbf{k}_{0})\sqrt{|K(\mathbf{k}_{0})|}}\bigg(\frac{1}{2\pi \rho}\bigg)^{(d-1)/2},
\end{equation}

plus terms of order $O\big(\rho^{(d-3)/2}\big)$. We can arrive at the same conclusion by locally parametrising the surface and using the classical stationary phase formula in Euclidean space. The result in the main text is obtained by relating the Gaussian curvature to the effective mass tensor as follows. The Gaussian curvature at a point $\mathbf{k} \in S$ can be obtained through the formula~\cite{curvature_adj}
\begin{eqnarray}
|K(\mathbf{k})| &= \frac{|\mathbf{v}(\mathbf{k}) \cdot \text{Adj}[\mathbf{H}(\mathbf{k})]\cdot\mathbf{v}(\mathbf{k})|}{v^{d+1}(\mathbf{k})}\nonumber\\
& = \frac{|\mathbf{v}(\mathbf{k})\cdot \mathbf{m}(\mathbf{k})\cdot\mathbf{v}(\mathbf{k})|}{|\mathbf{m}(\mathbf{k})|v^{d+1}(\mathbf{k})}
\end{eqnarray}

where $\text{Adj}$ denotes matrix adjugate with the identity $\text{Adj}[\mathbf{H}] = |\mathbf{H}|\mathbf{H}^{-1}$, and we introduce $\mathbf{m}(\mathbf{k})$, the inverse of the effective mass tensor of excitations in the periodic media, given by
\begin{equation}
	[\mathbf{m}^{-1}]_{ij} = \frac{\partial^2 \omega(\mathbf{k})}{\partial k_{i} \partial k_{j}}.
\end{equation}

$|\mathbf{m}(\mathbf{k})|$ then denotes the determinant. The vanishing of the curvature can be seen as the vanishing of the effective mass for both impulse and response in the direction of the group velocity of wavepackets centered at $\mathbf{k},$ or as effective mass diverging for responses in the transverse plane.

\section{\label{app:numerics} Numerical calculation of the exact Green's function}
The strongly singular manifold of the Green's function integrand presents significant calculational difficultly, so that to obtain exact results either recurrence relations \cite{horiguchi} or a small imaginary part $0i \to \epsilon i$ are typically used. However, the former suffers from instabilities for large $\bm{\rho},$ and needs to be derived for each lattice geometry, whilst the latter suffers an increase in computational cost due to the divergence of the integrand near $S$. To navigate around these difficulties we apply a higher-dimensional variant of integration by parts that gives us a regular integrand for $\Omega$ (with only divergent derivative). The integrand then permits standard (albiet expensive) quadrature techniques, whilst $\Gamma$ is obtained through the surface integral on the resonant level set. In one dimension repeated integration by parts on periodic functions $f, \omega$ gives

\begin{eqnarray}\label{appeq:ibp1}
   &\mathcal{P}\int \frac{dk f(k)}{\Delta - \omega(k)} = -\mathcal{P}\int dk\log|\Delta - \omega(k)| \Big(\frac{f(k)}{v(k)}\Big)' \nonumber\\
   &= \mathcal{P}\int dk(\Delta - \omega(k))\big[\log|\Delta - \omega(k)| - 1\big]
        \bigg(\frac{1}{v(k)}\Big(\frac{
    f'(k)}{v(k)}\Big)'\bigg)'\nonumber\\
    &,
\end{eqnarray}

when $v$ is non-vanishing on the support of $f.$ The prime denotes derivative. We note that endpoint contributions cancel via assumed periodicity, and the log divergences at either side of singularities can also be shown to cancel. If $v$ vanishes on the support of $f,$ we can decompose $f(k) = f(k)\phi_{\epsilon}(\Delta - \omega(k)) + f(k)\big(1 - \phi_{\epsilon}(\Delta - \omega(k))\big)$ for the bump function $\phi_{\epsilon}$, such that $\epsilon$ is small enough that the support of the first term does not contain any points of vanishing group velocity (i.e. as satisfied by the tubular region). The first term is then amenable to integration by parts via \eqref{appeq:ibp1}, and the second term is a regular integrand. In higher dimensions we accordingly make use of the divergence theorem $\int_{S}\nabla \cdot \mathbf{F} d\mathbf{k} = \int_{\partial S}\mathbf{F}\cdot d\mathbf{S},$ noting again that periodic boundary conditions and divergences of opposite signs cancel on boundaries to see
\begin{equation}\label{appeq:ibp2}
        \mathcal{P}\int \frac{d\mathbf{k} f(\mathbf{k})}{\Delta - \omega(\mathbf{k})} = -\mathcal{P}\int d\mathbf{k}\log|\Delta - \omega(\mathbf{k})|\nabla \cdot \Big(\frac{\mathbf{v}(\mathbf{k})f(\mathbf{k})}{v^2(\mathbf{k})}\Big),
\end{equation}
and applying integration by parts again for the final result. The bump function may be used here to regularize if necessary.

\section{Asymptotics beyond the critical angle\label{app:critical-angle}}

    We approximate the exponential decay rate of $\Gamma$ close to but beyond the critical angle corresponding to the caustic using saddle point analysis. Near a Van Hove singularity or when the dispersion varies rapidly, significant contributions to $\Omega$ may come from regions away from the resonant level set. We perform calculations in two dimensions for simplicity, but note that concepts may be extended to higher dimensions. We begin with 
    \begin{equation}
        \label{appeq:gam}{\Gamma}  = \mathcal{A} \int_{S}\frac{d\mathbf{k}}{(2\pi )}\frac{\psi(\mathbf{k},\mathbf{r})\psi^{*}(\mathbf{k},\mathbf{r'})}{{v}(\mathbf{k})} = \mathcal{A} \int_{S}\frac{d\mathbf{k}}{(2\pi )}\frac{\Phi(\mathbf{k})e^{i\mathbf{k} \cdot \bm{\rho}}}{{v}(\mathbf{k})},
    \end{equation}
    and suppose that some separation vector $\bm{\rho}_{\infty}$ lies exactly on a caustic corresponding to the wavevector $\mathbf{k}_{\infty}.$ If we then consider a unit length direction $\hat{\bm{\rho}} = (\hat{\bm{\rho}}_{\infty} + \hat{\bm{\epsilon}})$ just beyond the caustic with $\epsilon = |\hat{\bm{\epsilon}}| \ll 1$, we have
    \begin{eqnarray}
        &\hat{\bm{\epsilon}} \cdot \mathbf{v}(\mathbf{k}_{\infty}) = 0, \\
        &\bm{\rho}_{\infty} \parallel \mathbf{v}(\mathbf{k}_{\infty}).
    \end{eqnarray}
    That is, $\bm{\rho}_{\infty}$ is parallel to the caustic, and $\hat{\bm{\epsilon}}$ travels orthogonally beyond the caustic. The caustic is assumed global such that there is no travelling wave with real vector $\mathbf{k}$ such that $\hat{\bm{\rho}} \parallel \mathbf{v}(\mathbf{k})$ (such as the square lattice in the main text). Take a periodic parametrisation $\mathbf{k}(t)$ of $S$ with $\mathbf{k}_{\infty} = \mathbf{k}(t_{\infty})$, and consider the extension for $t$ in the complex plane, with small imaginary part. We seek the stationary point
    \begin{equation}
    (\hat{\bm{\rho}} + \hat{\bm{\epsilon}}) \cdot \mathbf{k}'(t) = 0. 
    \end{equation}

    For small $\epsilon,$ we expect the stationary point to comprise a correspondingly small correcting complex part $t'$ in $t$ and thus $\mathbf{k}.$ We expand 
\begin{eqnarray}
  &0 = \hat{\bm{\rho}} \cdot \mathbf{k}'(t) = \hat{\bm{\rho}} \cdot \mathbf{k}'(t_{\infty} + t') \nonumber\\
  &= (\hat{\bm{\rho}}_{\infty} + \hat{\bm{\epsilon}}) \cdot \left[\mathbf{k}'(t_{\infty}) + \mathbf{k}''(t_{\infty}) t' + \frac{t'^2}{2} \mathbf{k}'''(t_{\infty}) \right].
\end{eqnarray}
    To lowest order in ${\epsilon}$ and $t',$ we obtain 
\begin{equation}
        0 = \hat{\bm{\epsilon}} \cdot \mathbf{k}'(t_{\infty}) + \frac{t'^2}{2}  \hat{\bm{\rho}}_{\infty} \cdot \mathbf{k}'''(t_{\infty}). 
\end{equation}
    
Assuming the third derivative is non-zero (i.e., the derivative of the curvature is non-vanishing as we traverse the curve) the above equation is balanced for purely imaginary correction $t'$:  $i\sqrt{\epsilon} \sim t'$. The normalized argument of the exponent at the new stationary point has a real and imaginary part to leading order
    \begin{eqnarray}
    &\mathfrak{R}\big[ i\hat{\bm{\rho}} \cdot \mathbf{k}(t_{\infty} + t')\big] \sim -\epsilon^{3/2}, \\
    &\mathfrak{I}\big[i\hat{\bm{\rho}} \cdot \mathbf{k}(t_{\infty} + t')\big] \sim \hat{\bm{\rho}} \cdot \mathbf{k}(t_{\infty}).
    \end{eqnarray}

Now that the saddle point is obtained, we proceed to deform the contour of integration. As $\mathbf{k}$ is periodic with respect to $t$, the same period is respected when extended into the complex plane, in a small strip around the $t$ axis. We may then shift the contour into the complex plane where the boundary contributions cancel due to periodicity when the endpoints have the same imaginary part. We deform to a contour passing through the new stationary point with the real part of the exponent constant and scaling as $-\epsilon^{3/2}.$ Stationary phase then applies to the rapidly varying imaginary part, producing the usual inverse square root contribution in two dimensions. The difference is now that a small real part is contained in the exponential, and we obtain the long-range behaviour:
    \begin{equation}
    \Gamma \sim \frac{\exp\big[-\epsilon^{3/2}\rho \big]}{\sqrt{\rho}}.
    \end{equation}
    
    In two dimensions in particular, a small rotation of $\bm{\rho}_{\infty}$ by an angle $\epsilon$ corresponds to a perpendicular shift by $\bm{\epsilon}$ to leading order, so that the $\epsilon^{3/2}$ scaling of decay rate is also observed with variance of angle, as in the main text.
    In higher dimensions, one may heuristically extend the above line of reasoning and consider the higher dimensional analogue of steepest descent to arrive at a scaling in the generic case:
    \begin{equation}
    \Gamma \sim \frac{\exp\big[-\epsilon^{3/2}\rho \big]}{\rho^{(d-1)/2}}.
    \end{equation}
However, the geometry of saddle points is far more complex in higher dimensions, so that when the saddle point is degenerate, or multiple $\bm{\epsilon}$ are available, the scaling of both $\rho$ and $\epsilon$ can change, and the above is limited to a heuristic argument when $d>2$. We additionally note that the above analysis is limited to small perturbations, as inherent branch points of the parametrisation may dominate if the perturbation beyond the caustic is too large.

\input{Dispersion.bbl}

\end{document}

%% file: Dispersion.bbl
%